\documentclass[prd,showpacs,showkeys,floatfix,nofootinbib,
               preprint,12pt,tightenlines,fleqn]{revtex4}

\usepackage{amsmath,amssymb,graphicx,dcolumn}

\newcommand  {\version}{v5}        

\newcommand{\half}{{\textstyle \frac{1}{2}}}    
\newcommand{\dd}{\mathrm{d}}                    
\newcommand{\id}{\mathrm{d}}                    
\newcommand{\ii}{\mathrm{i}}                    
\newcommand{\beq}{\begin{equation}}
\newcommand{\eeq}{\end{equation}}
\newcommand{\beqa}{\begin{eqnarray}}
\newcommand{\eeqa}{\end{eqnarray}}

\begin{document}

\noindent Phys. Rev. D 78, 085026 (2008) \hfill arXiv:0809.321 [hep-ph] (\version)\newline\vspace*{0.5cm}
\title[New two-sided bound on isotropic Lorentz violation...]
      {\mbox{New two-sided bound on the isotropic Lorentz-violating parameter}
      of modified Maxwell theory\vspace*{.125\baselineskip}}
\author{F.R.~Klinkhamer}
\email{frans.klinkhamer@physik.uni-karlsruhe.de}
\author{M. Schreck}\email{schreck@particle.uni-karlsruhe.de}
\affiliation{\mbox{Institute for Theoretical Physics, University of Karlsruhe (TH),}\\
76128 Karlsruhe, Germany}

\begin{abstract}
\vspace*{2.5mm}\noindent
There is a unique Lorentz-violating
modification of the Maxwell theory of photons, which
maintains gauge invariance, CPT, and renormalizability.
Restricting the modified-Maxwell theory to the isotropic sector
and adding a standard spin--$\half$ Dirac particle $p^\pm$
with minimal coupling to the nonstandard photon $\widetilde{\gamma}$,
the resulting modified-quantum-electrodynamics model
involves a single dimensionless ``deformation parameter,''
$\widetilde{\kappa}_\text{tr}$.
The exact tree-level decay rates for two processes have been calculated:
vacuum Cherenkov radiation $p^\pm \to p^\pm\,\widetilde{\gamma}$
for the case of positive $\widetilde{\kappa}_\text{tr}$
and photon decay $\widetilde{\gamma} \to p^+\,p^-$
for the case of negative $\widetilde{\kappa}_\text{tr}$.
From the inferred absence of these decays for a particular high-quality
ultrahigh-energy-cosmic-ray event detected at the Pierre Auger Observatory
and a well-established excess of TeV gamma-ray events observed by the
High Energy Stereoscopic System telescopes,
a two-sided bound on $\widetilde{\kappa}_\text{tr}$
is obtained, which improves by eight orders of magnitude upon the best
direct laboratory bound.
The implications of this result are briefly discussed.
 \vspace*{0cm}\newline
\end{abstract}

\maketitle

\section{Introduction}
\label{sec:introduction}

Spacetime is a dynamic entity and quantum mechanics inescapably leads
to randomness. Combined, this suggests~\cite{Wheeler1957} that space may not
be perfectly smooth at very small length scales and that
Lorentz invariance may be fundamentally violated.

Precision tests of Lorentz invariance are, therefore,
of paramount importance and the ideal testing ground is the photonic sector.
If space, on the whole, is homogeneous and isotropic, there is
essentially one dimensionless parameter that describes
Lorentz violation in the photonic sector
(in a first approximation, this nonstandard photon is taken to be coupled
to standard Lorentz-invariant charged particles such as the electron).

The aim of the present article is to obtain a new bound on
this isotropic Lorentz-violating parameter.
Needless to say, the parameter may also have an entirely different
origin than the nontrivial spacetime structure mentioned above.
For the purpose of obtaining the bound, we remain ``agnostic'' as to the
potential origin of this particular Lorentz-violating parameter.

\section{Isotropic modified-Maxwell theory and current bounds}
\label{sec:isotropicmodMaxw-theory -current-bounds}

In this article, we consider an isotropic Lorentz-violating (LV)
deformation of quantum electrodynamics (QED)~\cite{JauchRohrlich1976}
given by the following action:
\beq\label{eq:modQED-action}
\hspace*{0mm}
S_\text{modQED}[\widetilde{\kappa}_\text{tr},e,M] =
S_\text{modMaxwell}[\widetilde{\kappa}_\text{tr}] +
S_\text{Dirac}[e,M] \,,
\eeq
with a modified-Maxwell term~\cite{ChadhaNielsen1983,ColladayKostelecky1998,
KosteleckyLanePickering2002,KosteleckyMewes2002,BaileyKostelecky2004}
for the gauge field $A_\mu(x)$ and the standard Dirac term~\cite{JauchRohrlich1976}
for the spinor field $\psi(x)$ relevant
to a spin--$\half$ particle (electric charge $e$ and mass $M$) and its
antiparticle (opposite charge and equal mass):
\begin{subequations}\label{eq:modM-standD-actions-ansaetze}
\beqa
S_\text{modMaxwell}[\widetilde{\kappa}_\text{tr}] &=& \int_{\mathbb{R}^4} \id^4 x \;
\Big( -\frac{1}{4}\,
\big[\, \eta^{\mu\rho}\eta^{\nu\sigma} +\kappa^{\mu\nu\rho\sigma}\,\big]
\, F_{\mu\nu}(x) \,F_{\rho\sigma}(x)\Big)\,,
\label{eq:modM-action}\\[2mm]
S_\text{Dirac}[e,M] &=&\int_{\mathbb{R}^4} \id^4 x \; \overline\psi(x) \Big(
\gamma^\mu \big(\ii\,\partial_\mu -e A_\mu(x) \big) -M\Big) \psi(x)\,,
\label{eq:standD-action}
\\[2mm]
F_{\mu\nu}(x) &\equiv& \partial_\mu A_\nu(x) - \partial_\nu A_\mu(x)\,,
\label{eq:field-strength}
\\[2mm]
\hspace*{-0mm}
\kappa^{\mu\nu\rho\sigma} &\equiv&
\frac{1}{2} \big(\,
\eta^{\mu\rho}\,\widetilde{\kappa}^{\nu\sigma} -
\eta^{\mu\sigma}\,\widetilde{\kappa}^{\nu\rho} +
\eta^{\nu\sigma}\,\widetilde{\kappa}^{\mu\rho} -
\eta^{\nu\rho}\,\widetilde{\kappa}^{\mu\sigma}  \,\big) \,,
\label{eq:nonbirefringent-ansatz}\\[2mm]
\hspace*{-0mm}
\big(\widetilde{\kappa}^{\mu\nu}\big) &\equiv&
\frac{3}{2}\;\widetilde{\kappa}_\text{tr} \;
\text{diag}\big(1,\,1/3,\,1/3,\,1/3\big)\,,
\label{eq:isotropic-ansatz}
\eeqa
\end{subequations}
where condition \eqref{eq:nonbirefringent-ansatz} on
the dimensionless deformation parameters $\kappa^{\mu\nu\rho\sigma}$
restricts the theory to the nonbirefringent
sector and the further condition \eqref{eq:isotropic-ansatz}
to the isotropic sector.
Here, and in the following, natural units are used with $c=\hbar=1$.
The fundamental constant $c$ now corresponds to the maximum attainable
velocity of the Dirac particle or, more importantly, to the
causal velocity from the underlying Minkowski spacetime with
Cartesian coordinates
$(x^\mu)$ $=$ $(x^0,\boldsymbol{x})$ $=$ $(c\,t,x^1,x^2,x^3)$
and metric $g_{\mu\nu}(x)$ $=$ $\eta_{\mu\nu}$ $\equiv$
$\text{diag}\,(+1$, $-1$, $-1$, $-1)\,$.

The particular \emph{Ans\"{a}tze}
\eqref{eq:nonbirefringent-ansatz}--\eqref{eq:isotropic-ansatz}
reduce the number of independent parameters in the
real background tensor $\kappa^{\mu\nu\rho\sigma}$
from 19 to 1. Physically~\cite{KosteleckyMewes2002}, the single remaining LV
parameter $\widetilde{\kappa}_\text{tr}$ changes the phase velocity of light to
$\sqrt{(1-\widetilde{\kappa}_\text{tr})/(1+\widetilde{\kappa}_\text{tr})}\,c$,
as will become clear
from \eqref{eq:definitions-xi-A-B}--\eqref{eq:disp-relations} below.

This deformation parameter $\widetilde{\kappa}_\text{tr}$ is,
however, difficult to measure.
The first bound by the Ives--Stilwell experiment~\cite{IvesStilwell1938}
was at the $1$ percent level
(see, e.g., Ref.~\cite{Tobar-etal2005} for a general discussion of this
type of experiment). At present, the best direct laboratory
bound~\cite{Reinhardt-etal2007} at the two--$\sigma$ level is
\beq\label{eq:lab-bound}
|\widetilde{\kappa}_\text{tr}| < 2 \times 10^{-7}\,,
\eeq
as follows from the penultimate unnumbered equation
in Ref.~\cite{Reinhardt-etal2007}.
The difficulty of bounding $\widetilde{\kappa}_\text{tr}$
in laboratory experiments is that its effects always appear in combination
with a quadratic Doppler factor $(\text{``}v\text{''}/c)^2$, where the
quotation marks point to the subtle issue of identifying the relevant
velocities~\cite[(b)]{Tobar-etal2005}.

A precision measurement~\cite{Odom-etal2006} of the electron anomalous
magnetic moment $a_e \equiv (g_e - 2)/2$,
combined with a four-loop calculation of standard QED~\cite{Gabrielse-etal2006},
gives an indirect bound~\cite{CaroneSherVanderhaeghen2006}
on $\widetilde{\kappa}_\text{tr}$ which is approximately a factor 6 better
than the direct bound \eqref{eq:lab-bound}.

Recently, an entirely different type of
indirect bound~\cite{KlinkhamerRisse2008a,KlinkhamerRisse2008b} has
been obtained from an ultrahigh-energy-cosmic-ray (UHECR) event,
which, at the two--$\sigma$ level, is given by
\beq\label{eq:UHECR-bound_MR2008b}
\widetilde{\kappa}_\text{tr} < 1.4 \times 10^{-19}\,.
\eeq
Clearly, this bound improves upon the laboratory
bound \eqref{eq:lab-bound} by many orders of magnitude,
but bound \eqref{eq:UHECR-bound_MR2008b}, as
it stands, is only one-sided. The main goal of the present article
is to obtain a two-sided bound.

\section{Lorentz-noninvariant decay processes}
\label{sec:Lorentz-noninvariant-decay-processes}

The modified--QED model defined by
\eqref{eq:modQED-action}--\eqref{eq:modM-standD-actions-ansaetze}
is gauge invariant, CPT invariant, and power-counting renormalizable
(renormalizability has been verified
at the one-loop level~\cite{KosteleckyLanePickering2002}).
The violation of Lorentz invariance leads to
modified propagation properties of the photon~$\widetilde{\gamma}$,
where a nonstandard symbol is used to emphasize its unusual properties.
The modified photon propagation, in turn, allows for
new types of particle decays.

\begin{figure*}[t]
\vspace*{0cm}
\begin{center}
\includegraphics[width=8cm]{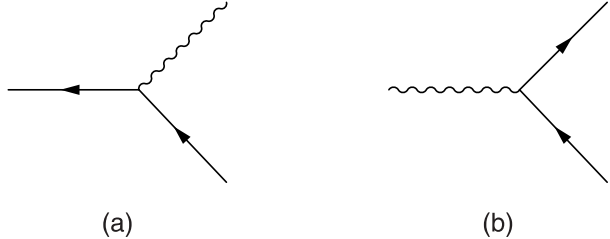} 
\end{center}
\vspace*{-0.5cm}
\caption{Feynman diagrams with time running from left to right
for (a) the vacuum-Cherenkov-radiation process and (b) the photon-decay process,
both evaluated for the isotropic Lorentz-violating
deformation of quantum electrodynamics given by
\eqref{eq:modQED-action}--\eqref{eq:modM-standD-actions-ansaetze}.}
\label{fig:FDcherenkov-decay}
\end{figure*}

In this article, we consider two such decay processes
(Figs.~\ref{fig:FDcherenkov-decay}ab),
whose occurrence depends on the sign of the LV parameter
in \eqref{eq:modQED-action}--\eqref{eq:modM-standD-actions-ansaetze}:
\begin{subequations}\label{eq:decay-a-b}
\beqa
\widetilde{\kappa}_\text{tr} > 0 &:&\quad  p^\pm \to p^\pm\,\widetilde{\gamma}  \,,
\label{eq:decay-a}\\[2mm]
\widetilde{\kappa}_\text{tr} < 0 &:&\quad  \widetilde{\gamma} \to p^{-}\,p^{+} \,,
\label{eq:decay-b}
\eeqa
\end{subequations}
where $p^-/p^+$ stands for the electron/positron particle
($e^-/e^+$ in standard notation) of the vectorlike $U(1)$ gauge theory
considered, that is, pure QED~\cite{JauchRohrlich1976}.
It is also possible to take the charged particles $p^+/p^-$ in
\eqref{eq:decay-a-b} to correspond to a simplified version of the
proton/antiproton (namely, a Dirac particle with partonic
effects neglected in first approximation).
Process \eqref{eq:decay-a} has been called ``vacuum Cherenkov radiation''
in the literature and process \eqref{eq:decay-b} ``photon decay.''
See Ref.~\cite{JacobsonLiberatiMattingly2006} for a general review
of LV decay processes
and Ref.~\cite{KaufholdKlinkhamer2006} for a detailed discussion
starting from Lorentz-noninvariant scalar models.

For both processes \eqref{eq:decay-a} and \eqref{eq:decay-b}
in the modified--QED model
\eqref{eq:modQED-action}--\eqref{eq:modM-standD-actions-ansaetze},
we have calculated the exact tree-level decay rates
and, in particular, the threshold energies
$E_\text{thresh}^{(a,b)}$. This work builds on earlier calculations
of Ref.~\cite{KaufholdKlinkhamer2007},
which also contains an extensive
list of references on standard and nonstandard Cherenkov radiation.
Certain technical details of the
calculation can be found in Ref.~\cite{KaufholdKlinkhamerSchreck2007}
and some of the results presented here have already been mentioned
in a recent review article~\cite{Klinkhamer2008}.

In order to write the results compactly, define
\begin{equation}\label{eq:definitions-xi-A-B}
\xi\equiv 2\,\widetilde{\kappa}_\text{tr}\,,\quad
\mathcal{A}\equiv 1/ \mathcal{B} \equiv \sqrt{(2+\xi)/(2-\xi)}\,.
\end{equation}
Precisely this combination $\mathcal{B}$  enters the photon dispersion
relation (leaving out the tilde on the suffix $\gamma$, for clarity):
\beq\label{eq:disp-relations}
E_\gamma(\vec{k})=\mathcal{B}\,|\vec{k}|\,,\quad
E_p(\vec{k})=\sqrt{|\vec{k}|^2+M^2}\,,
\eeq
where, for completeness, also the standard dispersion relation
of the charged Dirac particle $p$ has been given.

For the vacuum-Cherenkov-radiation process \eqref{eq:decay-a}
at $\xi>0$ with $E$ standing for the energy $E_p$
of the initial charged Dirac particle
(Fig.~\ref{fig:FDcherenkov-decay}a), the exact tree-level result
for the radiated-energy rate is
\beqa
\hspace*{-15mm}&&
\frac{\dd W^\text{(a)}}{\dd t}\,
\Bigg|_{E \geq E_\text{thresh}^\text{(a)}}^\text{exact}
=\,\frac{e^2}{4\pi}\;\,\frac{1}{3\,\xi^3 \,E\,\sqrt{E^2-M^2}}\;
\left(\mathcal{B}\:E-\sqrt{E^2-M^2}\right)^2
\nonumber\\[1mm]
\hspace*{-15mm}
&&\times \Bigg\{ 2\Big(\xi^2+4\,\xi+6\Big)E^2-\big(2+\xi\big)
\left(3\,\big(1+\xi\big)M^2+2\,\big(3+2\,\xi\big)\,
\mathcal{B}\,E\,\sqrt{E^2-M^2}\right)\!\Bigg\}\,,
\label{eq:dWdt-exact-positive-widetilde-kappa-tr}
\eeqa
where, from now on, the suffix `exact' on a rate abbreviates `exact tree-level.'
The threshold energy corresponds to the highest-energy zero of the above
expression and is given by
\begin{subequations}\label{eq:Ethreshold-positive-widetilde-kappa-tr}
\beqa
E_\text{thresh}^\text{(a)}
&=& M\; \sqrt{\big(1+\xi/2\big)/\xi}
\label{eq:Ethreshold-exact-positive-widetilde-kappa-tr}\\[1mm]
&=& M/\sqrt{\xi}+\mathsf{O}\big(M\,\sqrt{\xi}\big)\:,
\label{eq:Ethreshold-series-positive-widetilde-kappa-tr}
\eeqa
\end{subequations}
where the expansion holds for sufficiently small but positive $\xi$.
The threshold kinematics has the final fermion carrying
the full three-momentum of the initial fermion.

The energy behavior of the radiated-energy
rate \eqref{eq:dWdt-exact-positive-widetilde-kappa-tr} is not quite obvious and
the following two expansions at fixed
Dirac-particle mass $M$ and fixed LV parameter $0 < \xi \ll 1$ may be helpful:
\begin{subequations}
\label{eq:dWdt-near-threshold-asymptotic-positive-widetilde-kappa-tr}
\beqa
\hspace*{-15mm}&&
\frac{\dd W^\text{(a)}}{\dd t}\,\Bigg|_{E \geq E_\text{thresh}^\text{(a)}}
\approx \frac{e^2}{4\pi}\,\left(E-E_\mathrm{thresh}^\text{(a)}\right)^2\,
\left\{\frac{4}{3}\,\xi\, \left[\left(E/E_\mathrm{thresh}^\text{(a)}-1\right)
-\left(E/E_\mathrm{thresh}^\text{(a)}-1\right)^2\right.\right.
\notag \\[1mm]
\hspace*{-15mm}&&\,\hspace{3.25cm}\left.\left.
+\,\mathsf{O}\left(E/E_\mathrm{thresh}^\text{(a)}-1\right)^3\right]
+\,\mathsf{O}\left(\xi^2\,\left(E/E_\mathrm{thresh}^\text{(a)}-1\right)\right)\right\}
\label{eq:dWdt-near-threshold-positive-widetilde-kappa-tr}
\\[1mm]
\hspace*{-15mm}&&
\approx \frac{e^2}{4\pi}\,E^2
\Bigg\{\!\!\left(\frac{7}{24}\,\xi-\frac{1}{16}\,\xi^2+\mathsf{O}(\xi^3)\right)
+ \left(\! -1+\frac{1}{48}\,\xi-\frac{3}{32}\,\xi^2+
\mathsf{O}(\xi^3)\right)\frac{M^2}{E^2}+
\mathsf{O}\left(\frac{M^4}{\xi\,E^4}\right)\!\!\Bigg\},
\label{eq:dWdt-asymptotic-positive-widetilde-kappa-tr}
\eeqa
\end{subequations}
where the first expression holds for energies $E$ sufficiently close to but
above the threshold energy
[using the exact result \eqref{eq:Ethreshold-exact-positive-widetilde-kappa-tr}
throughout] and the second expression holds
for sufficiently large energies $E$.\footnote{The
approximation sign `$\approx$'
has been used in \eqref{eq:dWdt-near-threshold-asymptotic-positive-widetilde-kappa-tr}
because the remaining $\xi$ dependence is indicated only symbolically, not exactly.}
The corresponding expressions for the decay rate $\Gamma^{(a)}$ are
relegated to Appendix~\ref{app:vacuum-Cherenkov-decay-rate}.

For the photon-decay process \eqref{eq:decay-b} at $\xi<0$
with $E$ standing for the energy $E_\gamma$ of the initial photon
(Fig.~\ref{fig:FDcherenkov-decay}b),
the exact tree-level result for the decay rate is
\beqa
\Gamma^{(b)}\,\Big|_{E \geq E_\text{thresh}^\text{(b)}}^\text{exact}
&=&
\frac{e^2}{4\pi}\; \frac{2-\xi}{6\,(2+\xi)^{7/2}\,\sqrt{\left(1-\xi/2\right)}\,E^2}
\nonumber\\[1mm]
&&
\times \Big\{\mathcal{C}_4\,\big(\mathcal{C}_1-\mathcal{C}_2\big)
               +2\,\mathcal{A}\,E\,\mathcal{C}_3\,\big(\mathcal{C}_1+\mathcal{C}_2\big)
               +\mathcal{C}_5\,\big(\mathcal{C}_1^2-\mathcal{C}_2^2\big)\Big\}\,,
\label{eq:Gamma-exact-negative-widetilde-kappa-tr}
\eeqa
with
\begin{subequations}
\beqa
\mathcal{C}_1&\equiv&
\sqrt{\frac{(2+\xi)^2\,M^2+
\mathcal{A}\,E\,\big(2\,\mathcal{A}\,\xi\,E-\mathcal{C}_3\big)}{\xi\,(2+\xi)}}\,,
\\[1mm]
\mathcal{C}_2&\equiv&
\sqrt{\frac{(2+\xi)^2\,M^2+
\mathcal{A}\,E\,\big(2\,\mathcal{A}\,\xi\,E+\mathcal{C}_3\big)}{\xi\,(2+\xi)}}\,,
 \\[1mm]
\mathcal{C}_3&\equiv&
\sqrt{\xi\,(4-\xi^2)\,\big[\,2\,(2+\xi)\,M^2+\mathcal{A}^2\,\xi\, E^2\,\big]}\,,
\\[1mm]
\mathcal{C}_4&\equiv&
16\,\big[\,(1+\xi)\,M^2-\mathcal{A}^2\,\xi\, E^2\,\big]+4\,\xi^2\,M^2\,,
 \\[1mm]
\mathcal{C}_5&\equiv& 3\sqrt{2}\,\xi \,(2+\xi)\,E\,.
\eeqa
\end{subequations}
The corresponding  threshold energy is
\begin{subequations}\label{eq:Ethreshold-negative-widetilde-kappa-tr}
\beqa
E_\text{thresh}^\text{(b)}
&=&2\, M\;\sqrt{\big(1-\xi/2\big)/\big(-\xi\big)}
\label{eq:Ethreshold-exact-negative-widetilde-kappa-tr}\\[2mm]
&=&2\, M/\sqrt{\big(-\xi\big)}+\mathsf{O}\big(M\,\sqrt{-\xi}\big)\,,
\label{eq:Ethreshold-series-negative-widetilde-kappa-tr}
\eeqa
\end{subequations}
where the exact result \eqref{eq:Ethreshold-exact-negative-widetilde-kappa-tr}
has the same basic structure
as \eqref{eq:Ethreshold-exact-positive-widetilde-kappa-tr}
and the expansion in \eqref{eq:Ethreshold-series-negative-widetilde-kappa-tr}
holds for sufficiently small $|\xi|$.
The threshold kinematics has the three-momentum of the initial photon
split equally over the two final fermions.

The energy behavior of \eqref{eq:Gamma-exact-negative-widetilde-kappa-tr}
is clarified by the following two expansions at fixed
$M$ and fixed parameter $0 < -\xi \ll 1$:
\begin{subequations}
\beqa
\Gamma^{(b)}\,\Big|_{E \geq E_\text{thresh}^\text{(b)}}
&\approx&  \frac{e^2}{4\pi}\,
\sqrt{E_{\mathrm{thresh}}^\text{(b)}\,\left(E-E_{\mathrm{thresh}}^\text{(b)}\right)}
\,\left\{\Big(-\xi/\sqrt{2}\,\Big)\,
\left[ 1-\frac{5}{12}\left(E/E_\mathrm{thresh}^\text{(b)}-1\right)\right.\right.
\nonumber \\[1mm]
&&
+\left.\left.\,\mathsf{O}\left(\left(E/E_\mathrm{thresh}^\text{(b)}-1\right)^2\right)\right]
+\mathsf{O}\left(\xi^3\right)\right\}
\label{eq:Gamma-near-threshold-negative-widetilde-kappa-tr}
\\[1mm]
&\approx&
\frac{e^2}{4\pi}\;E\;\Bigg\{\left(-\frac{1}{3}\,\xi+\mathsf{O}(\xi^3)\right)
  +\mathsf{O}\left(\frac{M^4}{\xi E^4}\right)\Bigg\}\,,
\label{eq:Gamma-asymptotic-negative-widetilde-kappa-tr}
\eeqa
\end{subequations}
with the first expression holding for energies $E$ just above threshold
and the second for sufficiently large energies $E$.

Recalling the definitions \eqref{eq:definitions-xi-A-B} of
$\mathcal{B}$ and $\mathcal{A}$, the explicit presence of these
square roots in the rates \eqref{eq:dWdt-exact-positive-widetilde-kappa-tr}
and \eqref{eq:Gamma-exact-negative-widetilde-kappa-tr}
makes clear that, purely theoretically, the LV parameter
$\widetilde{\kappa}_\text{tr}$ cannot have an absolute value larger than 1.
(See, e.g., Refs.~\cite{AdamKlinkhamer2001,KosteleckyLehnert2001,BrosEpstein2002}
for a general discussion of causality,
unitarity, and stability in Lorentz-violating theories.)
But how small $|\widetilde{\kappa}_\text{tr}|$ really is,
is up to experiment to decide.

As a final technical remark, we repeat that the rates
calculated in this article apply to Dirac point particles.
This may be reasonable for photon decay into an electron-positron pair
but the results are only indicative for processes involving
hadrons (see also related remarks in, e.g., Ref.~\cite{Klinkhamer2008}
and Appendix~\ref{app:bounds-universal-isotropic-LV-parameter} of the
present article).

\section{New indirect bounds on the isotropic LV parameter}
\label{sec:new-bounds}

In this section, we turn to the experimental data
in order to obtain bounds on $\widetilde{\kappa}_\text{tr}$.
For positive $\widetilde{\kappa}_\text{tr}$,
the argument~\cite{Beall1970,ColemanGlashow1997}
is well-known (see also the review~\cite{Klinkhamer2008}
for further discussion):
from a single observed cosmic-ray event with primary energy $E_\text{prim}$,
the inferred absence of vacuum Cherenkov radiation \eqref{eq:decay-a}
implies $E_\text{prim}<E_\text{thresh}^\text{(a)}$, which,
with expression \eqref{eq:Ethreshold-positive-widetilde-kappa-tr},
gives an upper bound on $\widetilde{\kappa}_\text{tr}$.
For negative $\widetilde{\kappa}_\text{tr}$, the argument is similar:
from a single observed gamma-ray event with initial photon energy $E_\gamma$,
the inferred absence of photon decay \eqref{eq:decay-b}
implies $E_\gamma<E_\text{thresh}^\text{(b)}$ which,
with expression \eqref{eq:Ethreshold-negative-widetilde-kappa-tr},
gives a lower bound on $\widetilde{\kappa}_\text{tr}$.

For positive $\widetilde{\kappa}_\text{tr}$, this is indeed how the
previously mentioned bound \eqref{eq:UHECR-bound_MR2008b} was obtained
from the detection at the Pierre Auger Observatory~\cite{Auger2004}
of an $E_\text{prim}=148\;\text{EeV}$ UHECR event~\cite{Abraham-etal2008}.
Here, we use another Auger event~\cite{Abraham-etal2006,Barbosa-etal2004}
with a slightly higher energy,
$E_\text{prim}=212\;\text{EeV}=2.12 \times 10^{20}\;\text{eV}$.
This  particular Auger event is of the so-called hybrid type~\cite{Auger2004},
which means that it has been observed by both surface
detectors and fluorescence telescopes;
see Table~\ref{tab:exp-data} for further details.
Demanding $E_\text{prim} < E_\text{thresh}^{(a)}$ and
using \eqref{eq:Ethreshold-positive-widetilde-kappa-tr}
with $M$ set to the value $52\;\text{GeV}$ of an iron nucleus
(an absolutely conservative choice as discussed in
Refs.~\cite{KlinkhamerRisse2008a,KlinkhamerRisse2008b}),
the following two--$\sigma$ ($98\,\%\;\text{CL}$) bound is obtained:
\begin{subequations}
\beq\label{eq:UHECR-upperbound}
\widetilde{\kappa}_\text{tr} < 6 \times 10^{-20}\,,
\eeq
which is lower than \eqref{eq:UHECR-bound_MR2008b}, mainly
because of the larger primary energy used.
For completeness, the three--$\sigma$ ($99.9\,\%\;\text{CL}$) bound is
$\widetilde{\kappa}_\text{tr} < 8 \times 10^{-20}$.

For negative $\widetilde{\kappa}_\text{tr}$, we use the detection by
the High Energy Stereoscopic System (HESS) imaging atmospheric
Cherenkov telescopes~\cite{HESS2003}
of $E_\gamma \geq 30\;\text{TeV}$ photons from the shell-type supernova remnant
\mbox{RX J1713.7--3946} \cite{Aharonian-etal2006,Aharonian-etal2007}.
These gamma-ray photons with energies above $30\;\text{TeV}$
have been detected at the five--$\sigma$ level~\cite{Aharonian-etal2007}.
For definiteness, we consider a fiducial photon event
with $\overline{E}_\gamma=30\;\text{TeV}=3.0 \times 10^{13}\;\text{eV}$ and
a relative error in the individual reconstructed energy
of the order of $15\,\%$ as quoted in Ref.~\cite{Aharonian-etal2006};
see Table~\ref{tab:exp-data} for further details.
Demanding $\overline{E}_\gamma < E_\text{thresh}^{(b)}$ and
using \eqref{eq:Ethreshold-negative-widetilde-kappa-tr}
with $M$ set to the value $511\;\text{keV}$ of an electron,
the following two--$\sigma$ ($98\,\%\;\text{CL}$) bound is obtained:
\beq\label{eq:HESS-lowerbound}
-\widetilde{\kappa}_\text{tr} < 9 \times 10^{-16}\,.
\eeq
\end{subequations}
For completeness, the three--$\sigma$ ($99.9\,\%\;\text{CL}$) bound is
$-\widetilde{\kappa}_\text{tr} < 1.1 \times 10^{-15}$.

Related bounds in a modified-QED model
with additional isotropic Lorentz violation in the fermionic sector
are given in Appendix~\ref{app:bounds-in-another-modified-QED-model}.

At this moment, it may be appropriate to mention that
the orders of magnitude of bounds \eqref{eq:UHECR-upperbound}
and \eqref{eq:HESS-lowerbound} [the first one scaled to $M=M_\text{proton}$;
see below] agree with the qualitative
limits of Coleman and Glashow~\cite{ColemanGlashow1997}.
The improvement, here, is that the exact tree-level decay rates have been
calculated (and found to be nonvanishing)
and that a proper error analysis has been performed.
Still, the fact remains that the qualitative analysis of
Ref.~\cite{ColemanGlashow1997} has been confirmed remarkably well
by the quantitative results of the present article.

\begin{table*}
\begin{center}
\caption{Experimental results from the Pierre Auger Observatory~\cite{Auger2004}
and the HESS Cherenkov-telescope array~\cite{HESS2003}
for the energy values (with relative errors for the two
individual events) used for bounds
\protect\eqref{eq:UHECR-upperbound} and \protect\eqref{eq:HESS-lowerbound}.
The value of the Auger-event energy~\cite{Abraham-etal2006} has been increased
by $5\%$ to account for the missing energy~\cite{Barbosa-etal2004}
of a hadronic primary
(the energy value given in Ref.~\cite{Abraham-etal2006} corresponds to a
hypothetical photon primary). For this hybrid Auger event,
also the shower-maximum atmospheric depth $X_\text{max}$
has been measured (the measured $X_\text{max}$ value rules out a
photonic primary at the three--$\sigma$ level~\protect\cite{Abraham-etal2006}).
The target position of the HESS events has values for the
right ascension $\alpha$ and declination $\delta$ that correspond to those
of the supernova remnant
\mbox{RX J1713.7--3946}~\protect\cite{Aharonian-etal2006,Aharonian-etal2007}.
\vspace*{0mm}}
\label{tab:exp-data}
\renewcommand{\tabcolsep}{0.25pc}   
\renewcommand{\arraystretch}{1.1}   
\begin{tabular}{lcc|ccr}
\hline\hline
experiment
& observation
& Ref.
& energy $E$
& $\Delta E/E$
& additional characteristics
 \\
\hline
Auger
& ID 737165
& \protect\cite{Abraham-etal2006}
& $E_\text{prim}=212 \;\text{EeV}$   
& $25\,\%$
& $X_\text{max} = 821\;\text{g}\:\text{cm}^{-2}$
\\
HESS
& 2003--2005
& \protect\cite{Aharonian-etal2007}
& $\overline{E}_\gamma=30\;\text{TeV}$
& $15\,\%$
& $\big( \alpha,\,\delta \big)$ $=$
  $\big( 17^\text{\,h}\,13^\text{\,m}\,33^\text{\,s},\,
         -39^{\,\circ}\,45^{\,\prime}\,44^{\,\prime\prime}\big)$
\\
\hline\hline
\end{tabular}
\end{center}
\end{table*}

Three last remarks are as follows.
First, the above threshold bounds only make
sense if the travel length of the decaying particle is less than the
source distance. In general terms, this was already noticed in
Ref.~\cite{ColemanGlashow1997}.
With the specific results of Sec.~\ref{sec:Lorentz-noninvariant-decay-processes}
and the methodology of Ref.~\cite{KaufholdKlinkhamer2006}, it indeed
follows that the average travel lengths $L^{(a)}$ and $L^{(b)}$
from the processes \eqref{eq:decay-a} and \eqref{eq:decay-b},
for the $\widetilde{\kappa}_\text{tr}$ and $M$ values considered in this article,
are of the order of meters (or even less) rather than the parsec-scale distances
of the astronomical sources involved ($D \sim 1\;\text{Mpc}$
for the case of UHECRs~\cite{Abraham-etal2008}
and $D \sim 1\;\text{kpc}$
for the case of supernova remnant \mbox{RX J1713.7--3946}~\cite{Aharonian-etal2006}).

Second, the events from
Table~\ref{tab:exp-data} used for the above bounds are far from unique.
Similar events have also been seen by other experiments,
for example, the UHECR event with $E_\text{prim}=320\;\text{EeV}$
and $X_\text{max} = 815\;\text{g}\:\text{cm}^{-2}$
observed by the Fly's Eye detector~\cite{Bird-etal1995}
and the $E_\gamma=8\;\text{TeV}$ gamma-rays from the galaxy M87
detected by VERITAS~\cite{Acciari-etal2008}.

Third,  the right-hand side of bound \eqref{eq:UHECR-upperbound}
or \eqref{eq:HESS-lowerbound} scales as $(M/E)^2$,
with $E$ the energy of the incoming particle and $M$
the mass of the charged Dirac particle in the respective process
\eqref{eq:decay-a} or \eqref{eq:decay-b}.
This functional dependence shows the potential for improvement
if higher energy events become available in the
future\footnote{Bound \eqref{eq:HESS-lowerbound}, for example,
can be reduced by a factor $7$ if the 1.5--$\sigma$
detection at HESS~\cite{Aharonian-etal2007} of
$80\;\text{TeV}$ gamma-ray photons from \mbox{RX J1713.7--3946}
is confirmed by further observations.}
and, for the UHECR case, if light hadronic primaries can be
selected.\footnote{Bound \eqref{eq:UHECR-upperbound}
can be reduced by a factor $12$ if a mass value
$M= 15\;\text{GeV}$ is used, which corresponds to the mass of an oxygen nucleus.
For the single UHECR event of Table~\ref{tab:exp-data},
this particular (conservative)  mass value
is suggested by the measured $X_\text{max}$ value (cf. Sec. III of
Ref.~\cite{KlinkhamerRisse2008a}). Still, a proper analysis will require
more events and perhaps other diagnostics in addition to $X_\text{max}$.}

\section{Conclusion}
\label{sec:conclusion}

From the results of the previous section, the final two-sided two--$\sigma$
bound on the single isotropic Lorentz-violating parameter of modified--QED theory
\eqref{eq:modQED-action}--\eqref{eq:modM-standD-actions-ansaetze} is
\beq\label{eq:UHECR-two-sided-bound}
- 9 \times 10^{-16} < \widetilde{\kappa}_\text{tr} < 6 \times 10^{-20}\,.
\eeq
The simple model \eqref{eq:modQED-action}--\eqref{eq:modM-standD-actions-ansaetze}
can be considered to be embedded in a
Standard Model Extension~\cite{ColladayKostelecky1998}
with all physical (renormalized) Lorentz-violating parameters set to zero,
except for one, namely, the isotropic, CPT--even, Lorentz-violating parameter
for the dimension-four term with two Maxwell field strength tensors of the
photon field.\footnote{With nonzero weak mixing angle~\cite{PeskinSchroeder1995},
the $W^\pm$ and $Z^0$ vector fields also display modified propagation
(and interaction) properties, because of $SU(2)\times U(1)$ gauge invariance
[the action is given
by \eqref{eq:modSM-action}--\eqref{eq:isotropic-ansatz-kappatilde-i}
in Appendix~\ref{app:bounds-universal-isotropic-LV-parameter} with
$\overline{\kappa}_\text{tr}^{(1)} =\overline{\kappa}_\text{tr}^{(2)}
 \equiv\widetilde{\kappa}_\text{tr}$ and $\overline{\kappa}_\text{tr}^{(3)}=0$].
However, these weak-vector-boson modifications do not directly
affect the tree-level calculations of processes \eqref{eq:decay-a} and
\eqref{eq:decay-b}, with the fermions considered as Dirac point particles.}

If the same isotropic Lorentz-violating parameter
applies equally to all gauge bosons of the Standard Model,
bound \eqref{eq:UHECR-two-sided-bound} can be replaced by a stronger one,
which has a lower bound at $-2 \times 10^{-19}$
(see Appendix~\ref{app:bounds-universal-isotropic-LV-parameter}).
But the derivation of this particular lower bound involves certain
assumptions on the parton distributions
(the error is, however, still dominated by the
relatively large error from the primary-energy measurement).
For the moment, bound \eqref{eq:UHECR-two-sided-bound} for the purely photonic
parameter is to be preferred, as it does not involve a partonic calculation.

This new indirect bound \eqref{eq:UHECR-two-sided-bound} improves
by 8 orders of magnitude upon the direct laboratory bound \eqref{eq:lab-bound}.
Observe that, strictly speaking, \eqref{eq:UHECR-two-sided-bound} is also a
``terrestrial bound,'' as it relies on having detected a primary
traveling over a few hundred meters in the Earth's atmosphere,
the precise astronomical origin of the primary being of secondary importance
(see the first of the last three remarks in Sec.~\ref{sec:new-bounds}).
For the type of experiments involved
(Auger~\cite{Auger2004} and HESS~\cite{HESS2003}),
the Earth's atmosphere is an integral part of the instrument.

Combined with previous astrophysics bounds~\cite{KosteleckyMewes2002}
on the 10 birefringent modified-Maxwell-theory parameters
at the $10^{-32}$ level
and ``terrestrial'' UHECR bounds~\cite{KlinkhamerRisse2008b} on the 8
nonisotropic nonbirefringent parameters at the $10^{-18}$ level,
bound \eqref{eq:UHECR-two-sided-bound}
gives the following two--$\sigma$
bound on \emph{all} entries of the background tensor $\kappa^{\mu\nu\rho\sigma}$
in the general modified-Maxwell action \eqref{eq:modM-action}:
\beq\label{eq:UHECR-kappa-tensor-bound}
\max_{\,\{\mu,\nu,\rho,\sigma\}} \; |\kappa^{\mu\nu\rho\sigma}|< 5\times 10^{-16} \,,
\eeq
where the fact has been used that the largest entry of
$|\kappa^{\mu\nu\rho\sigma}|$ has a value
$(1/2)\,|\widetilde{\kappa}_\text{tr}|$ if the other 18 parameters
are negligibly small. For completeness, the three--$\sigma$
bound is $\max\,|\kappa^{\mu\nu\rho\sigma}| < 6 \times 10^{-16}$.
Remark that bounds \eqref{eq:UHECR-two-sided-bound}
and \eqref{eq:UHECR-kappa-tensor-bound}
hold in a Sun-centered, nonrotating frame of reference.

In order to put \eqref{eq:UHECR-kappa-tensor-bound} in perspective, recall
that the equality of gravitational and inertial mass was experimentally
verified by  Newton (1687), Bessel (1832), and E\"{o}tv\"{o}s (1889) at the
$10^{-3}$, $10^{-5}$, and $10^{-7}$ levels,
respectively~\cite{FischbachTalmadge1999}.
These experiments then led Einstein to the Principle of Equivalence,
the cornerstone of the General Theory of Relativity~\cite{Einstein1916}.
In our case, the conclusion from \eqref{eq:UHECR-kappa-tensor-bound}
would be (elaborating on remarks in
Refs.~\cite{Collins-etal2004,GagnonMoore2004,BernadotteKlinkhamer2007})
that the local Lorentz invariance left-over from
a fundamental theory of quantum spacetime is exact
and that the underlying principle for this apparent fact
remains to be determined.

\section*{ACKNOWLEDGMENTS}  

It is a pleasure to thank W. Hofmann, V.A. Kosteleck\'{y}, and M. Risse
for helpful discussions.

\section*{NOTE ADDED}

Recently, an indirect bound on $|\widetilde{\kappa}_\text{tr}|$
at the $10^{-11}$ level was obtained from particle-collider
data~\cite{Hohensee-etal2008}, which improves by $4$ orders of
magnitude upon the direct laboratory bound \eqref{eq:lab-bound}.

\appendix
\section{vacuum--Cherenkov decay rate}
\label{app:vacuum-Cherenkov-decay-rate}

In this appendix, the decay rate for the vacuum-Cherenkov process
$p^{\pm}$ $\to$ $p^{\pm}\,\widetilde{\gamma}$ is given, where
$p^{\pm}$ stands for a Dirac point particle with charge $\pm e$ and mass $M$.
The theory is defined by
\eqref{eq:modQED-action}--\eqref{eq:modM-standD-actions-ansaetze} and
this particular decay process occurs for the case of a positive
Lorentz-violating parameter $\xi\equiv 2\,\widetilde{\kappa}_\text{tr}$
(the relevant Feynman diagram appears in Fig.~\ref{fig:FDcherenkov-decay}a).

With $E$ denoting the energy $E_p$ of the initial charged Dirac particle,
the exact tree-level result for the decay rate is
\begin{align}
\hspace*{-0mm}\Gamma^{(a)}\Big|_{E\geq E_{\mathrm{thresh}}^{(a)}}^\text{exact}
&=\frac{e^2}{4\pi}\;\frac{2}{3\,\xi^2\,\sqrt{4-\xi^2}\,E\,\sqrt{E^2-M^2}}\;
\left(\mathcal{B}\,E-\sqrt{E^2-M^2}\,\right) \notag \\[2mm]
\hspace*{-0mm}&
\quad\,
\times \Big\{-\left(3\,\xi^2+6\,\xi+8\right)\,E^2
+\left(3\,\xi^2+8\,\xi+4\right) M^2
\notag \\[2mm]&\qquad\quad\,
+\,\mathcal{B}\,\left(3\,\xi^2+10\,\xi+8\right)\,E\,\sqrt{E^2-M^2}\,\Big\}\,,
\label{eq:Gamma-exact-positive-widetilde-kappa-tr}
\end{align}
where $\mathcal{B}=\mathcal{B}(\xi)$ has been defined
in \eqref{eq:definitions-xi-A-B}.

 From \eqref{eq:Gamma-exact-positive-widetilde-kappa-tr},
the threshold energy is found to be the same as
\eqref{eq:Ethreshold-exact-positive-widetilde-kappa-tr}
obtained from the radiated-energy rate
\eqref{eq:dWdt-exact-positive-widetilde-kappa-tr}.
In fact, this threshold energy traces back to the common
factor $\left(\mathcal{B}\,E-\sqrt{E^2-M^2}\,\right)$
in \eqref{eq:Gamma-exact-positive-widetilde-kappa-tr}
and \eqref{eq:dWdt-exact-positive-widetilde-kappa-tr}.

The energy behavior of rate \eqref{eq:Gamma-exact-positive-widetilde-kappa-tr}
is clarified by the following two expansions at fixed
mass $M$ and fixed LV parameter $0 < \xi \ll 1$:
\begin{subequations}
\begin{align}
\hspace*{-5mm}\Gamma^{(a)}\Big|_{E\geq E_{\mathrm{thresh}}^{(a)}}
&\approx  \frac{e^2}{4\pi}\,\left(E-E_\mathrm{thresh}^\text{(a)}\right)\,
\left\{2\,\xi\, \left[ \left(E/E_\mathrm{thresh}^\text{(a)}-1\right)
-\frac{4}{3}\,\left(E/E_\mathrm{thresh}^\text{(a)}-1\right)^2
\right.\right.
\notag \\[1mm]
\hspace*{-15mm}&\,\hspace{0.25cm}\left.\left.
+\,\mathsf{O}\left(\left(E/E_\mathrm{thresh}^\text{(a)}-1\right)^3\right)\right]
+\,\mathsf{O}\left(\xi^2\,\left(E/E_\mathrm{thresh}^\text{(a)}-1\right)\right)\right\}
\label{eq:Gamma-near-treshold-positive-widetilde-kappa-tr}
\\[1mm]
&\approx \frac{e^2}{4\pi}\;E\;\Bigg\{
\left(\frac{2}{3}\,\xi-\frac{1}{24}\,\xi^2+\mathsf{O}(\xi^3)\right)
\notag\\[2mm]&\quad\,
+\left(-\frac{3}{2}-\frac{1}{6}\,\xi-\frac{23}{96}\,\xi^2+\mathsf{O}(\xi^3)\right)\frac{M^2}{E^2}
+\mathsf{O}\left(\frac{M^4}{\xi E^4}\right)\Bigg\}\,,
\label{eq:Gamma-asymptotic-positive-widetilde-kappa-tr}
\end{align}
\end{subequations}
where the first expression holds for energies $E$ just above threshold
and the second expression for sufficiently large energies $E$.

\section{Bounds on isotropic Lorentz violation in another modified--QED model}
\label{app:bounds-in-another-modified-QED-model}

In this appendix, bounds are given on isotropic Lorentz violation
in a modified-QED model with additional isotropic $c$--type
Lorentz violation in the fermionic sector~\cite{ColladayKostelecky1998}.

For two fermion species, this modified--QED model has the following action:
\beq\label{eq:modQED2-action}
\hspace*{0mm}
S_\text{modQED2} =
S_\text{modMaxwell} +S_\text{modDirac2} \,,
\eeq
with the modified-Maxwell term given by \eqref{eq:modM-standD-actions-ansaetze}
and the modified--Dirac term by
\begin{subequations}
\beqa
\hspace*{-8mm}
S_\text{modDirac2} &=&\int_{\mathbb{R}^4} \id^4 x \; \sum_{j=1}^{2}\;
\overline\psi^{(j)}(x)
\Big( \big[\eta_{\mu\nu}+c^{(j)}_{\mu\nu}\,\big]\,
\gamma^\mu \big(\ii\,\partial^\nu -e^{(j)} A^\nu(x) \big) -M^{(j)}\Big) \psi^{(j)}(x)\,,
\label{eq:modDirac2-action}
\\[2mm]
\hspace*{-8mm}
\big(c^{(j)}_{\mu\nu}\,\big) &=&
c^{(j)}_{00}\;\text{diag}\big(1,\,1/3,\,1/3,\,1/3\big)\,,
\label{eq:isotropic-ansatz-cmunu}
\eeqa
\end{subequations}
where `$j$' in \eqref{eq:modDirac2-action} is the fermion species label.
The action \eqref{eq:modQED2-action} will be applied to an idealized
physical situation with only photons, electrons, and protons.

From the identification of appropriate coordinate
transformations~\cite{BaileyKostelecky2004}, it then follows that,
to leading order,
bounds \eqref{eq:UHECR-upperbound} and \eqref{eq:HESS-lowerbound}
are replaced by
\begin{subequations}
\beqa
 \Big[\widetilde{\kappa}_\text{tr}-(4/3)\,c_{00}^{(p)}\,\Big]&<& 6\times 10^{-20}\,,
\label{eq:UHECR-upperbound-fermionic}\\[2mm]
-\Big[\widetilde{\kappa}_\text{tr}-(4/3)\,c_{00}^{(e)}\,\Big]&<& 9\times 10^{-16}\,,
\label{eq:HESS-lowerbound-fermionic}
\eeqa
\end{subequations}
where the suffixes `$(p)$' and `$(e)$' refer to the possible isotropic $c$--type
Lorentz violation of the proton and electron, respectively.
Bound \eqref{eq:UHECR-upperbound-fermionic} is, most likely,
rather conservative, because it is based on the mass of an iron nucleus
(which contains both protons and neutrons),
as discussed in the second paragraph of Sec.~\ref{sec:new-bounds}.

\section{Bounds on a universal isotropic LV parameter}
\label{app:bounds-universal-isotropic-LV-parameter}

Consider a particular deformation~\cite{ColladayKostelecky1998} of the
Standard Model of elementary particles, where the same isotropic
Lorentz-violating parameter applies universally to all gauge bosons
and other Lorentz-violating parameters are absent (or, at least, negligible).
Specifically, the action is given by
\beq\label{eq:modSM-action}
\hspace*{0mm}
S_\text{modSM} =
S_\text{modYM} +S_\text{restSM}\,,
\eeq
with the only modification occurring in the kinetic terms of
the Lie-algebra-valued Yang--Mills gauge fields $A_\mu^{(i)}(x)$,
\begin{subequations}\label{eq:modYM-ansaetze-kappatilde-universal}
\beqa
\hspace*{-5mm}
S_\text{modYM} &=& \int_{\mathbb{R}^4} \id^4 x \;
\Bigg(\frac{1}{2}\,\sum_{i=1}^{3}\,\text{tr}\,
\Big( \big[ \eta^{\mu\rho}\eta^{\nu\sigma}
+\overline{\kappa}^{(i)\;\mu\nu\rho\sigma}\big]\,
F_{\mu\nu}^{(i)}(x)\, F_{\rho\sigma}^{(i)}(x)\Big)\Bigg)\,,
\label{eq:modYM-action}\\[2mm]
F_{\mu\nu}^{(i)}(x) &\equiv&
\partial_\mu A_\nu^{(i)}(x) - \partial_\nu A_\mu^{(i)}(x)
+ g^{(i)}\,\big[A_\mu^{(i)}(x),A_\nu^{(i)}(x)\big] \,,
\label{eq:YM-field-strength}\\[2mm]
\hspace*{-0mm}
\overline{\kappa}^{(i)\;\mu\nu\rho\sigma} &\equiv&
\frac{1}{2} \big(\,
\eta^{\mu\rho}\,\overline{\kappa}^{(i)\;\nu\sigma} -
\eta^{\mu\sigma}\,\overline{\kappa}^{(i)\;\nu\rho} +
\eta^{\nu\sigma}\,\overline{\kappa}^{(i)\;\mu\rho} -
\eta^{\nu\rho}\,\overline{\kappa}^{(i)\;\mu\sigma}  \,\big)\, ,
\label{eq:nonbirefringent-ansatz-kappatilde-universal}\\[2mm]
\hspace*{-0mm}
\big(\overline{\kappa}^{(i)\;\mu\nu}\big) &\equiv&
\frac{3}{2}\;\overline{\kappa}_\text{tr}^{(i)} \;
\text{diag}\big(1,\,1/3,\,1/3,\,1/3\big)\,,
\label{eq:isotropic-ansatz-kappatilde-i}\\[2mm]
\hspace*{-0mm}
\overline{\kappa}_\text{tr}^\text{univ} &\equiv&
 \overline{\kappa}_\text{tr}^{(1)}
=\overline{\kappa}_\text{tr}^{(2)}
=\overline{\kappa}_\text{tr}^{(3)}\,,
\label{eq:isotropic-ansatz-kappatilde-universal}
\eeqa
\end{subequations}
where `$\text{tr}$' in \eqref{eq:modYM-action} stands for the trace and
$F_{\mu\nu}^{(i)}$ in \eqref{eq:YM-field-strength} is the standard
(Lie-algebra-valued) Yang--Mills field strength~\cite{PeskinSchroeder1995},
with label $i=1$, $2$, and $3$ referring to
the gauge group $U(1)$, $SU(2)$, and $SU(3)$, respectively.
Since $S_\text{restSM}$ in \eqref{eq:modSM-action} is not written down explicitly,
there is no need to say more about the normalization of the anti-Hermitian Lie-algebra
generators, except that setting $\overline{\kappa}_\text{tr}^\text{univ}=0$
in \eqref{eq:modSM-action}--\eqref{eq:modYM-ansaetze-kappatilde-universal}
reproduces the Standard Model action;
 see Refs.~\cite{ColladayKostelecky1998,PeskinSchroeder1995} for details.
Note that such a \emph{universal} form of Lorentz violation could come
from a nontrivial small-scale structure of classical
spacetime over which the gauge bosons propagate
freely~\cite{BernadotteKlinkhamer2007}.\footnote{The particular calculation
of Ref.~\cite{BernadotteKlinkhamer2007} found spacetime ``defects''
to affect primarily gauge bosons, not fermions.
It remains to be seen if this conclusion holds generally.}

\begin{figure*}[t]
\vspace*{0cm}
\begin{center}
\includegraphics[width=4cm]{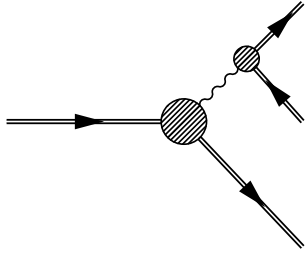}
\end{center}
\vspace*{-0.75cm}
\caption{Parton-model proton-break-up process \eqref{eq:decay-c}
for the isotropic Lorentz-violating Standard Model Extension
\eqref{eq:modSM-action}--\eqref{eq:modYM-ansaetze-kappatilde-universal}.}
\label{fig:FDcherenkov-proton-break-up}
\end{figure*}

A lower bound on $\overline{\kappa}_\text{tr}^\text{univ}$
can be obtained from the proton-break-up process involving an
electron-positron pair [the process will be labeled (c) in this appendix]:
\beq\label{eq:decay-c}
\overline{\xi}\equiv 2\,\overline{\kappa}_\text{tr}^\text{univ} < 0 :\quad
p^{+} \to p^{+}\,e^{-}\,e^{+}\,.
\eeq
This process can be calculated in the parton-model approximation
(Fig.~\ref{fig:FDcherenkov-proton-break-up})
and has already been considered in Ref.~\cite{GagnonMoore2004},
to which the reader is referred for further details
(for a general introduction to the parton model, see, e.g.,
Ref.~\cite{PeskinSchroeder1995}).
The corresponding energy threshold for $|\overline{\xi}|\ll 1$ is given by
\beq\label{eq:threshold-decay-c}
E_\text{thresh}^\text{(c)}=
M_p\,\big/\sqrt{\big(-\overline{\xi}\,\big)\big(\chi_p - \chi_e\big)} \,,
\eeq
where $M_p$ is the proton mass and
the numbers $\chi_{p,e} \in [0,1]$ characterize the total
$U(1)$, $SU(2)$, and $SU(3)$ gauge boson content of the proton
and the electron, respectively.

Two remarks on \eqref{eq:threshold-decay-c} may be helpful.
First, the precise meaning of
$\chi_p$ (or $\chi_e$) is that it is the sum of the first integral moments
of the relevant (gauge-boson) parton distribution functions
in the proton (or electron)~\cite{GagnonMoore2004}.
Second, the heuristics behind \eqref{eq:threshold-decay-c} is that,
with only gauge bosons
affected by Lorentz violation, the maximum speeds of proton and electron
differ, because of their different gauge-boson content.
These different speeds then allow for this particular proton-break-up process.
It must be emphasized that the process occurs only due to the nontrivial
parton content of proton and electron (remark that the analogous process
$p^{+} \to p^{+}\,p^{-}\,p^{+}$ still does not occur).

Now turn to \eqref{eq:threshold-decay-c} to get a bound on
the parameter $\overline{\kappa}_\text{tr}^\text{univ}$ by the same type of
argument as employed in Sec.~\ref{sec:new-bounds}, namely, that
the mere observation of a cosmic ray implies
$E_\text{prim} < E_\text{thresh}^\text{(c)}$, which then gives
the desired bound. For a proton $p$, neutron $n$, or electron $e$
of energy $E= 10^{2}\;\text{EeV}$,
one has $\chi_p - \chi_e \approx \chi_n - \chi_e \approx 0.35$,
according to Table 1 of Ref.~\cite{GagnonMoore2004}.
(The bulk of $\chi_{p,n} - \chi_e$ comes from the
different gluon content of nucleon and electron, namely,
$\chi_{p,n}^{(3)} - \chi_e^{(3)}  \approx 0.45$.)
Next, scale the proton mass  $M_{p}$ in \eqref{eq:threshold-decay-c} up to the
value $M_{\text{prim}} = 52\;\text{GeV}$
and use the $E_{\text{prim}} = 212\;\text{EeV}$ event of Table 1,
in order to get the following two--$\sigma$ bound:
\beq
- \overline{\kappa}_\text{tr}^\text{univ}< 2\times 10^{-19}\,.
\eeq

Combined with the previous bound \eqref{eq:UHECR-upperbound}
which also holds for the theory considered in this appendix,
the final two-sided two--$\sigma$  bound reads:
\beq\label{eq:UHECR-two-sided-bound-kappatilde-universal}
- 2\times 10^{-19}<\overline{\kappa}_\text{tr}^\text{univ}<6\times 10^{-20}\,,
\eeq
where, for the lower bound,
the relatively large error from the primary-energy measurement
dominates the theoretical uncertainties of the partonic calculations.


\end{document}